# Resolving Few-Layer Antimonene/Graphene Heterostructures

Tushar Gupta,[1] Kenan Elibol,[2] Stefan Hummel,[2,3] Michael Stöger-Pollach,[4] Clemens Mangler,[2] Gerlinde Habler,[5] Jannik C. Meyer,[2] Dominik Eder,[1,*] Bernhard C. Bayer[1, 2,*]

*1. Institute of Materials Chemistry, Vienna University of Technology (TU Wien), Getreidemarkt 9/165, A-1060 Vienna, Austria.*

*2. Faculty of Physics, University of Vienna, Boltzmanngasse 5, A-1090 Vienna, Austria*

*3. GETec Microscopy GmbH, Seestadtstrasse 27, A-1220 Vienna, Austria*

*4. USTEM, Vienna University of Technology (TU Wien), Wiedner Hauptstrasse 8-10, A-1040 Vienna, Austria*

*5. Department of Lithospheric Research, University of Vienna, Althanstrasse 14, A-1090 Vienna, Austria*

**Corresponding authors:* bernhard.bayer-skoff@tuwien.ac.at, dominik.eder@tuwien.ac.at

## Abstract

Two-dimensional (2D) antimony (Sb, "antimonene") recently attracted interest due to its peculiar electronic properties and its suitability as anode material in next generation batteries. Sb however exhibits a large polymorphic/allotropic structural diversity, which is also influenced by the Sb's support. Thus understanding Sb heterostructure formation is key in 2D Sb integration. Particularly 2D Sb/graphene interfaces are of prime importance as contacts in electronics and electrodes in batteries. We thus study here few-layered 2D Sb/graphene heterostructures by atomic-resolution (scanning) transmission electron microscopy. We find the co-existence of two Sb morphologies: First is a 2D growth morphology of layered β-Sb with β-Sb(001)||graphene(001) texture. Second are one-dimensional (1D) Sb nanowires which can be matched to β-Sb with β-Sb[2-21]⊥graphene(001) texture and are structurally also closely related to thermodynamically non-preferred cubic Sb(001)||graphene(001). Importantly, both Sb morphologies show rotational van-der-Waals epitaxy with the graphene support. Both Sb morphologies are well resilient against environmental bulk oxidation, although superficial Sb-oxide layer formation merits consideration, including formation of novel epitaxial $Sb_2O_3$(111)/β-Sb(001) heterostructures. Exact Sb growth behavior is sensitive on employed processing and substrate properties including, notably, the nature of the support underneath the direct graphene support. This introduces the substrate *underneath* a direct 2D support as a key parameter in 2D Sb heterostructure formation. Our work provides insights into the rich phase and epitaxy landscape in 2D Sb and 2D Sb/graphene heterostructures.

## Introduction

Among the two-dimensional (2D) pnictogens (i.e., group 15/VA elements, incl. P, As, Sb and Bi)[1–4] particularly mono- and few-layered 2D Sb ("antimonene") has recently attracted increasing research interest.[5,6] Firstly, this is due to 2D Sb's peculiar electronic properties towards novel 2D electronics incl. layer-dependent (semi-)metal-to-semiconductor transition[7,8] high carrier mobilities,[9] strain-tuneable indirect-to-direct band gap transition,[7,8] and the possibility of two- and three-dimensional topological insulator behavior for mono-[10] and few-layered[5] 2D Sb, respectively. Secondly, the recent high interest equally results from 2D Sb's high suitability for sustainable energy and catalysis applications, incl. as an anode material in next-generation Li- and Na-ion batteries[11–21] as well as in (electro-)catalysis,[22–24] supercapacitors,[25] charge extraction in photovoltaics,[26] and thermoelectrics.[27]

Sb however shows a large polymorphic/allotropic structural diversity. This includes several layered, potential 2D forms[8,28–31] such as in particular the thermodynamically most stable, rhombohedral, buckled honeycomb-structured β-Sb (A7, R-3m, 166)[22,23,30–44] and the metastable orthorhombic, puckered "washboard"-structured α-Sb (A17, cmca, 64).[31,45,46] Additionally several non-layered metastable allotropes at high pressure and in thin film form have been reported incl. simple cubic, body-centered-cubic, face-centered-cubic and hexagonally-close-packed Sb.[47–56] Most of these phases are related via small atomic rearrangements,[31,54–56] and some even have been suggested to show thickness dependent phase transitions in nanostructures.[31] This polymorphicity calls for close control over Sb's structure in any potential synthesis scenario for the various desired application fields. The structure of Sb deposits is however not only determined by kinetic growth process conditions but is also intimately linked to the Sb deposits' support, an effect which is exacerbated for ultrathin 2D Sb. Therefore, understanding Sb heterostructure formation is key for controllable 2D Sb growth. This is true not only for Sb's use in 2D electronics, where typically laterally large, defect-free 2D Sb films are desired,[33] but also for Sb's use in catalysis and energy applications, where often nano-sized 2D Sb deposits with a large number of edges are preferred.[22,23]

Among the various possible Sb heterostructures,[5] in particular 2D Sb/graphene interfaces are of prime importance for two reasons: First, in the context of catalysis and energy applications, Sb/carbon hybrids are emerging as a highly useful materials combination, e.g. in batteries,[11–20] electrocatalysis[22,23] and supercapacitors.[25] 2D Sb/graphene heterostructures can readily approximate such Sb/carbon hybrids in order to understand their as-of-yet little elucidated interface properties. Second, in the context of 2D electronics, recent work has suggested that device contacts formed by 2D Sb/graphene heterostructures could be technologically advantageous towards tuning contact resistances.[57,58]

The structural properties and formation mechanisms of Sb heterostructures incl. 2D Sb/graphene, remain however as-of-yet largely underexplored, in particular at the atomically resolved level. To address this, we provide here an atomic-scale (scanning) transmission electron microscopy ((S)TEM) investigation into the properties of a few-layered 2D Sb/graphene heterostructure model system that, as we find, readily emulates Sb/carbon heterostructures as manufactured by vapor phase techniques for electronics[38,48,50,51] and also as synthesized by wet chemistry routes for energy applications.[12,14,15,19,20] Our approach thereby facilitates direct assessment of interfacing and epitaxial effects in 2D Sb/carbon heterostructures, with graphene also acting as an ideal support[59] for the employed atomic resolution (S)TEM techniques.[60]

Our work reveals in our 2D Sb/graphene heterostructures, the co-existence of a 2D growth morphology of layered β-Sb(001)||graphene(001) phase and texture as well as of a one-dimensional (1D) Sb growth morphology. The latter 1D morphology can be matched to β-Sb with β-Sb[2-21]⊥graphene(001) texture but also to a non-layered, thermodynamically non-preferred cubic Sb(001)||graphene(001). Importantly, both Sb morphologies show preferred relative crystallographic (mis)orientations with respect to the supporting graphene monolayer lattice, indicating that rotational van-der-Waals (vdW) epitaxy can readily exist in 2D Sb/graphene heterostructures. Both Sb morphologies are found to be well resilient against environmental oxidation in ambient atmosphere although superficial surface oxidation is shown to be important to consider, particularly due to here suggested formation of epitaxial $Sb_2O_3$(111)/β-Sb(001) heterostructures. We find that exact Sb growth results are sensitive on employed processing techniques and substrate properties incl., notably, the nature of the support *underneath* the direct graphene support. Our work thereby provides fundamental insights into the rich phase and epitaxy relations in 2D Sb and 2D Sb/graphene heterostructures.

## Results and Discussions

**Morphology and Structure.** We first characterize in Fig. 1 the morphology and structure of the few layer Sb on graphene model system, which is prepared by physical vapor deposition (PVD) of Sb onto chemical vapor deposited (CVD) monolayer graphene. We first focus on optimized Sb deposition conditions towards high Sb crystallinity, with the wider parameter space of the Sb PVD on graphene being discussed further below. The nominal 10 nm thick Sb deposits in Figs. 1-4 were thermally evaporated onto monolayered CVD graphene films.[61,62] During Sb PVD (base pressure ~$10^{-5}$ mbar), the graphene substrates were held at room temperature (RT) and also at controlled temperatures of 150 °C and 250 °C. The graphene either remained on its Cu CVD catalyst foils[61,62] during Sb PVD (Fig. 1a,b, Fig. 2) or was additionally also transferred prior to Sb PVD to be suspended as a freestanding monolayer membrane across holey TEM grids[63] ((i.e. no Cu foils underneath, Fig. 1c-g, Figs. 3-4). Nominal deposited Sb thickness was measured via a co-exposed (non-heated) quartz crystal microbalance. After Sb deposition, samples were stored in ambient air. Further details on methods can be found in the Supporting Information.

The scanning electron microscopy (SEM) image of the 250 °C deposition in Fig. 1a reveals that under our optimized PVD conditions, the Sb deposits on the graphene form isolated islands with two distinctly different base shapes: First are flat 2D Sb deposits with (truncated) hexagonal or (truncated) triangular base shapes. Second are rod-like 1D Sb deposits with rectangular bases. Lateral extents of all Sb deposits are in the range of tens to hundreds of nm. While such lateral sizes are too small for device fabrication in 2D electronics, they are compatible with the requirements for 2D Sb catalysis and energy applications.[22] Importantly, such feature sizes also provide a large enough Sb/graphene heterostructure model system for convenient elucidation of Sb phases and interfacing with high resolution (S)TEM. Notably, as shown in Fig. 1a, the edges of both the triangular-/hexagonal-shaped and the rod-shaped Sb deposits show a high degree of visually apparent directional alignment amongst each phase type, respectively. This is a first indication of potential epitaxy effects between our Sb deposits and their graphene support and will be further examined below.

The Raman spectrum corresponding to the 250 °C deposition in Fig. 1b displays primarily two peaks at low wavenumbers that are characteristic for elemental Sb (117 $cm^{-1}$; 154 $cm^{-1}$). These peaks are best matched with $E_g$ and $A_{1g}$ modes of few layer β-Sb, respectively, but are also potentially consistent with α-Sb and/or pressure-induced phases of Sb.[8,31,33,46,54] We note that thicker Sb deposits may be overrepresented in Raman intensity.[42,43] Significant volume Sb-oxide formation can be excluded based on our Raman data as the signal intensity at wavenumbers corresponding to Sb-oxides is comparatively weak (e.g. for thermodynamically most stable $Sb_2O_3$ expected at ~190 $cm^{-1}$ and 250 $cm^{-1}$).[64] Raman peaks characteristic for graphene (G at 1593 $cm^{-1}$; and 2D at 2701 $cm^{-1}$) are also found in Fig. 1b, consistent with the high quality CVD graphene used as substrate.[61,62] The absence of a significant defect-related D-peak at ~1350 $cm^{-1}$ confirms that the CVD graphene support was not degraded during Sb PVD. Thereby our Raman data also confirms that no covalent Sb-carbon bond formation has occurred and that our 2D Sb/graphene interfaces are of vdW-type,[38] consistent with theoretical predictions.[8,57,58]

To assess the crystallographic structure of the Sb deposits in a localized fashion, we employ in Fig. 1c-g aberration-corrected, atomically-resolved and element-specific STEM (Nion UltraSTEM 100 at 60 kV electron acceleration voltage) in annular dark field (ADF)[60] mode to image individual Sb deposits at high resolution in top plan view. Corresponding ADF STEM and bright-field (BF) TEM data from focused-ion-beam (FIB) cross-sections in Fig. 2 provide a complementary side view of the Sb deposits. Supporting Figs. S1 and S2 provides atomic models and Fourier transform (FT)/selected area electron diffraction (SAED) simulations of all identified phases.

The first group of interest is the flat Sb deposits from Fig. 1a with (truncated) hexagonal (Fig. 1c) or triangular base shape (Fig. 1d). The phase identification for these structures is straightforward: At atomic resolution and view from top, all flat hexagonal/triangular deposits show a six-fold symmetric appearance that can be best indexed via the FT of their atomic resolution images to rhombohedral β-Sb viewed along the [001] zone axis (Fig. 1h) i.e. with the basal (001) layers of the layered 2D β-Sb parallel to the graphene(001) substrate (i.e. β-Sb[001]⊥graphene(001) = β-Sb(001)||graphene(001)). This phase identification to β-Sb(001) is also fully corroborated by the corresponding side view of a triangular/hexagonal deposits in Fig. 2b-d, which clearly resolves the layered nature of the β-Sb(001) when viewed along the [110] zone axis, with the β-Sb(001) planes parallel to the graphene(001) substrate. Delineating projected edge directions correspond to [100], [010] and [110] in the top view STEM images for both hexagonal and triangular β-Sb (Fig. 1c,d). Hexagonal and triangular deposits typically appear flat in STEM images, indicating (001) top surfaces. The observation of 2D β-Sb(001)||graphene(001) is in line with recent literature.[31–39]

The second group of interest are the rod-like 1D Sb deposits with rectangular bases (Fig. 1e-g). In top view at atomic resolution, these structures always show a FT with four-fold symmetry. Their phase identification is less straight forward: On the one hand, the STEM data agrees with β-Sb when viewed along the [2-21] zone axis i.e. at a texture of β-Sb[2-21]⊥graphene(001). Notably, β-Sb with [2-21] zone axis perpendicular to support does not have a defined low (hkl) value interface plane parallel to the support when viewed from the side, but only slightly inclined base planes (Fig. 1i; an approximation for an interface plane would be β-Sb(10 -10 23)). β-Sb[2-21] is closely related to AA-stacked α-Sb multilayers via a small shear deformation.[31] Recently, a thickness-dependent crossover from α-Sb to β-Sb[2-21] has been suggested to occur in 1D Sb deposits.[31] On the other hand, the fourfold symmetry STEM images of the 1D rod-like Sb deposits also matches well with a thermodynamically non-preferred simple cubic, non-layered Sb polymorph viewed along its [001] zone axis (Fig. 1j) i.e. cubic Sb[001]⊥graphene(001) = cubic Sb(001)||graphene(001).[47,50] Cubic Sb polymorphs are related to rhombohedral β-Sb via a unidirectional deformation.[54–56] While the existence of cubic Sb *in bulk form* has been a long-standing matter of debate in literature,[52–56] cubic Sb is typically associated with high pressure conditions but has also been reported to occur in Sb thin films, presumably formed via substrate-induced stress.[47–56] Notably, structurally β-Sb[2-21] (and AA α-Sb multilayers) and cubic Sb(001) structures are all closely related and may gradually transition into each other.[31,54–56] This makes their differentiation difficult and partly ambiguous. This is also underlined by, e.g., the cross-sectional TEM of a rod-like Sb deposit in Fig. 2d which shows lattice planes with a principal spacing of ~0.3 nm. These are consistent with β-Sb[2-21] as well as cubic Sb(001) viewed from the side (Fig. 1i,j). Therefore for the remainder of this report, we refer to the 1D Sb morphology as "β-Sb[2-21]/cubic Sb(001)" phase and texture, suggesting that the 1D rods are compatible with both β-Sb[2-21]⊥graphene(001) and cubic Sb(001)||graphene(001). For determination of β-Sb[2-21]/cubic Sb(001) in-plane epitaxial relations to graphene support, delineating projected edge directions and similar discussion below we will use the cubic Sb unit cell and associated (hkl) plane and [uvw] direction notation since this conveniently simplifies the description of the crystallographic system compared to the inclined plane β-Sb[2-21] description. Delineating projected edge directions in the top view STEM images are [110] for the rods (Fig. 1e-g). Fig. 1e-g and Fig. 2d show that the 1D morphology does not have a flat top surface but rather very strong faceting to a pyramidal shape over a square (Fig. 1e) or rectangular (Fig. 1f,g) base. To best reproduce the observed angles in Fig. 1e-g and Fig. 2d the delineating faceted surfaces have to be of (223) family in cubic Sb notation (Supporting Fig. S3). Alternatively, also (111) family facets can provide a reasonable match. 1D Sb morphologies with four-fold atomic symmetries that co-exist with 2D β-Sb(001) have been observed in older work on Sb/graphite[50,51] and also recent work on Sb/graphene,[31] albeit other recent work under very similar conditions for Sb/graphene heterostructures did not observe 1D nanostructure growth.[38] We note that one recent paper[65] ascribed 1D Sb nanostructures to β-Sb(001), albeit without providing direct crystalline structure confirmation for their assignment.

We have confirmed the morphology-structure relation of triangular/hexagonal base shape corresponding to layered 2D β-Sb(001) and rectangular rod base shape corresponding to β-Sb[2-21]/cubic Sb(001) via >60 atomic resolution STEM, lattice resolution TEM and SAED observations. This makes us confident that we can safely assign the crystallographic phase of a deposit via its macroscopic base shape as observed in lower magnification SEM or atomic force microscopy (AFM) data.[51]

Following this approach, AFM data (taken via conventional AFM as well as via correlated AFM-SEM, GETec AFSEM, see Supporting Fig. S4) indicates for depositions at 250 °C, for the layered 2D β-Sb(001) deposits, a minimum thickness of 4.7 nm (equivalent to ~12 layers,[8,29,42] i.e. few layer antimonene) and an average thickness (± standard deviation) of 21±14 nm. The 1D rod-like β-Sb[2-21]/cubic Sb(001) deposits are relatively thicker, with a minimum thickness of 10 nm and an average thickness of 31±10 nm. To estimate the relative abundance of 2D β-Sb(001) and 1D β-Sb[2-21]/cubic Sb(001) in our deposits, we compute from SEM and AFM images for the 250 °C depositions, both domain number counts and average equivalent feature sizes (see Supporting Information for calculation) for each phase: Via this analysis, we find a lower number fraction of 2D β-Sb(001) domains (40±1 count-%) compared to β-Sb[2-21]/cubic Sb(001) (60±1 count-%). These 2D β-Sb(001) domains grow to however cover a relatively larger area (60±5 area-% for 2D β-Sb(001) vs. 40±5 area-% for β-Sb[2-21]/cubic Sb(001)). The observation that the 2D β-Sb(001) islands grow to larger lateral sizes is also reflected in a larger maximum and average equivalent feature size for the 2D β-Sb(001) (maximum: 260 nm; average:

113±95 nm) compared to the β-Sb[2-21]/cubic Sb(001) (maximum: 130 nm; average: 72±54 nm).

Comparing our results with prior literature we note that overall morphology and size of our Sb domains on carbon substrates are consistent not only with vacuum-based vapor deposition techniques as usually used in electronics[38,48,50,51] (akin to our PVD synthesis) but also with several wet-chemistry synthesis routes (incl. using $SbCl_3$[12,14,19,20] and ball-milled and annealed Sb/carbon mixtures[15]) as usually used in energy materials synthesis. This highlights that our here investigated 2D Sb/graphene heterostructure model system is relevant to a wide range of synthesis conditions and electronics and energy-related application profiles of Sb on carbon.

In terms of application potential, we note that trigonally deformed Sb (like simple cubic Sb) has recently been predicted to feature superior thermoelectrical performance over β-Sb.[56] Given that monolayered 2D β-Sb has been predicted to surpass all other pristine 2D materials in terms of thermoelectric performance,[27] future studies on band structure and electronic properties of the here observed β-Sb[2-21]/cubic Sb(001) deposits merit consideration.

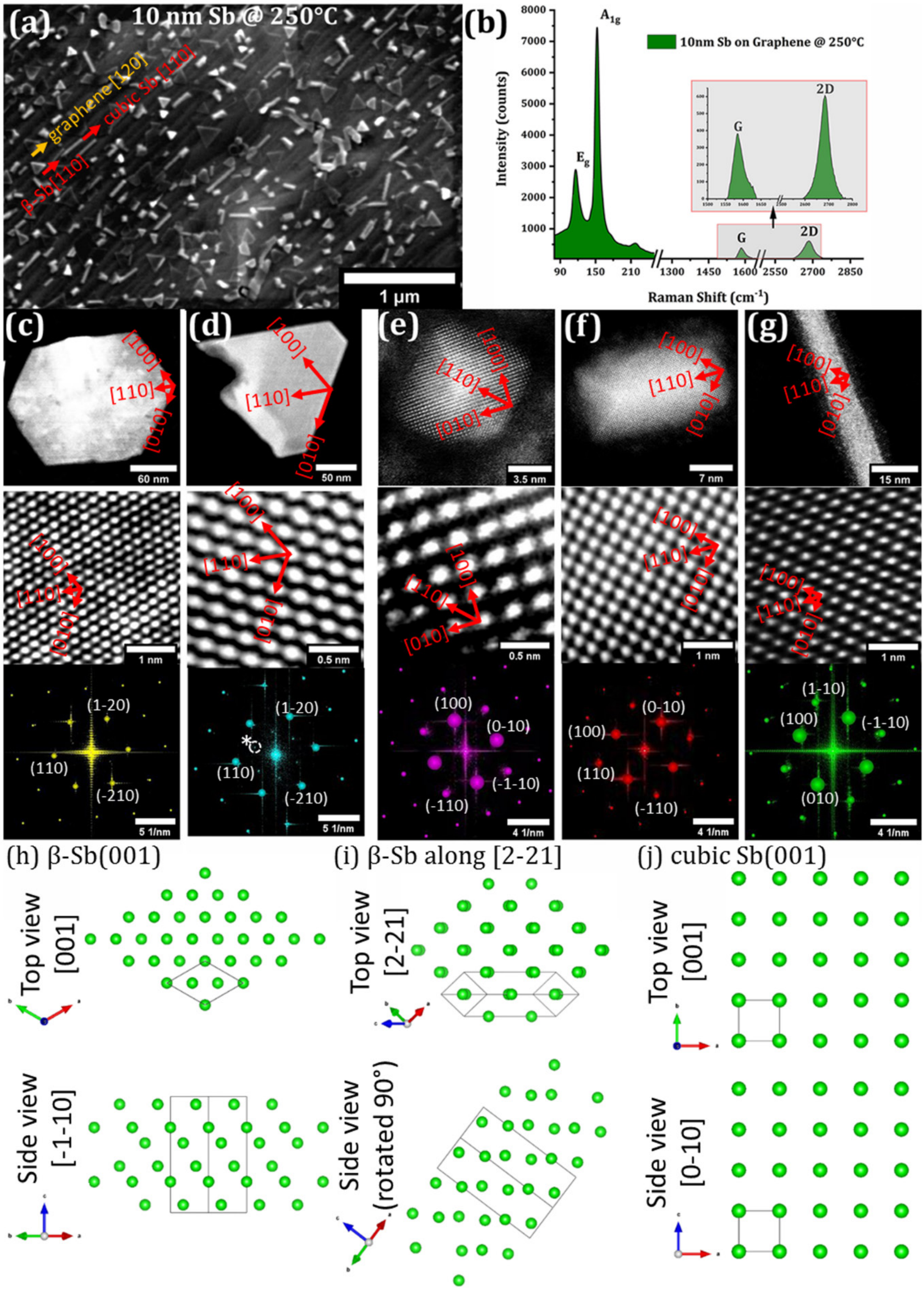

**Figure 1:** (a) SEM image and (b) Raman spectrum of 10 nm Sb deposited at 250 °C onto Cu-supported graphene. Salient lattice directions in the graphene and the Sb are labelled in (a), as identified in the main text. The minor peak in (b) at ~215 $cm^{-1}$ is related to minor Cu-oxide formation on bare regions of the Cu support during ambient air storage.[66,67] (c-

g) ADF STEM images of individual particles of 10 nm Sb deposited at 150°C and 250°C onto suspended monolayer graphene, showing overview (top) and atomic resolution images (middle) and corresponding FTs (bottom). The FTs are indexed to β-Sb(001) viewed along [001] zone axis (c,d) and cubic Sb(001) viewed along [001] zone axis (e,f,g), respectively. Corresponding salient crystallographic directions are superimposed over the ADF images. The six-fold symmetric "*"-indexed reflection set in (d) is ascribed cubic $Sb_2O_3$ viewed along the [111] zone axis and corresponds to $Sb_2O_3$ (2-20) reflection family, as described below. We note that under our STEM imaging conditions no electron beam induced phase transitions or materials modifications to the Sb deposits were observed.[68–70] (h-j) Atomic models of β-Sb(001), β-Sb[2-21] and cubic Sb(001), respectively. For further information on atomic models and FT simulations see Supporting Figs. S1 and S2.

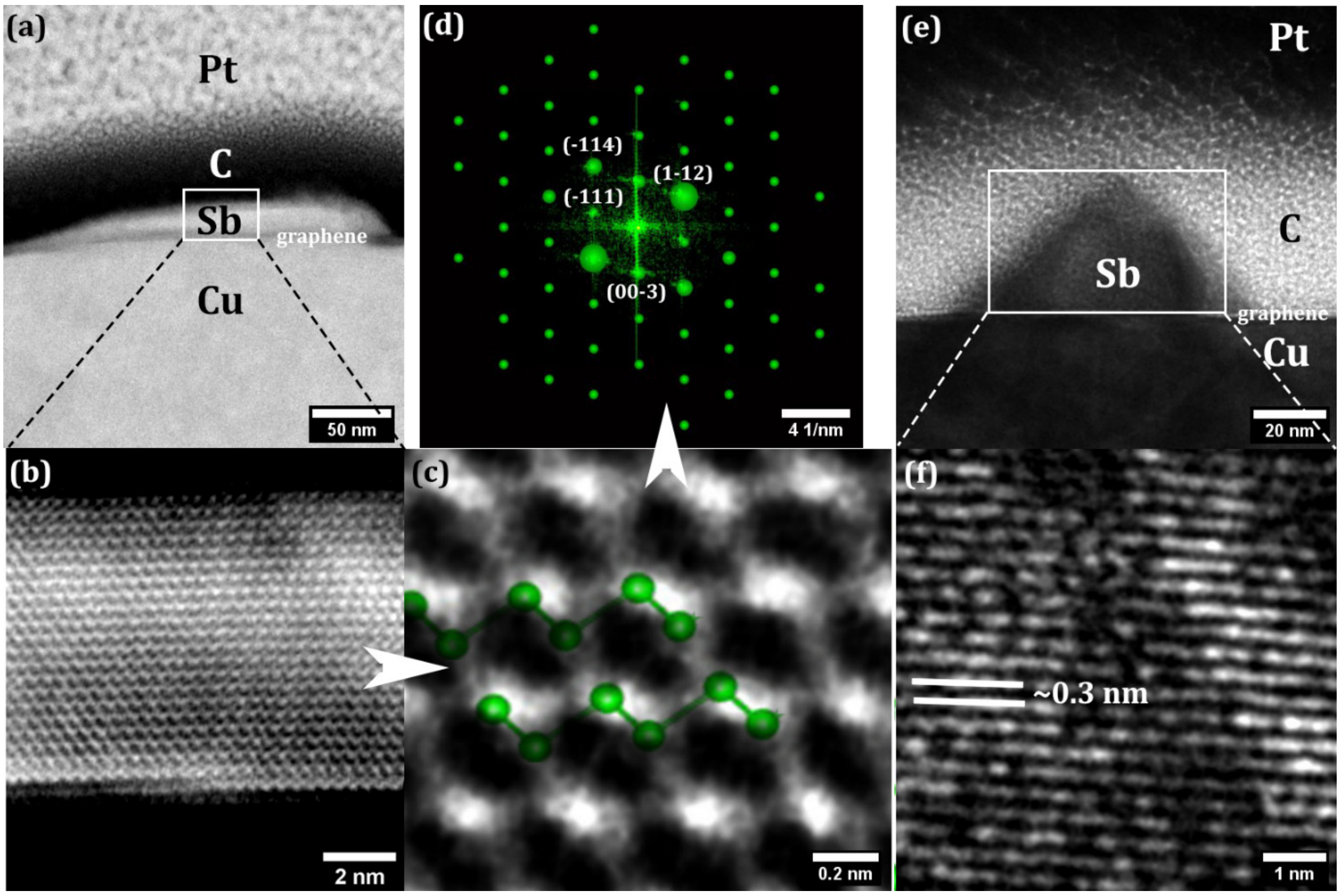

**Figure 2:** Cross-sectional STEM/TEM of 10 nm Sb deposited onto Cu-supported graphene: (a-c) show ADF STEM of a β-Sb(001) deposit in overview (a), intermediate (b) and high resolution (c). (d) shows the FT of (c) indexed to β-Sb(001) viewed along the [110] zone axis. In (c) a schematic of the β-Sb(001) layers is superimposed as illustration. (e,f) shows BF-TEM of a β-Sb[2-21]/cubic Sb(001) deposit in overview (e) and at high resolution (f).

**Van-der-Waals Epitaxy.** So far our data has shown that we have grown 2D Sb/graphene heterostructures, where the Sb deposits are comprised of two co-existing morphologies, namely few-layer 2D β-Sb(001) and 1D nanorods β-Sb[2-21]/cubic Sb(001). Importantly, for *both* these Sb morphologies Figs. 1a indicated a high degree of directional alignment of their respective domain edges on the monolayer graphene support. Given the vdW nature of the Sb/graphene interface (Fig. 1b), three mechanisms could contribute to such alignment: First is direct rotational vdW epitaxy between the growing Sb and its graphene support directly underneath.[71] For 2D β-Sb(001) direct epitaxial relationships with various substrates have been reported incl., e.g., $WSe_2$,[39] tellurides,[32,36] mica[33] and Ge.[35] Particularly, for β-Sb(001) on graphene prior work has given a mixed picture: Some work[38] reported rotational vdW epitaxy for β-Sb/graphene via indirect measurements, while other work observed no such epitaxy.[31,44,50,51] For the 1D nanorod β-Sb[2-21]/cubic Sb(001), epitaxial effects have to date not been reported.[31,51] Therefore, the question if *direct* vdW epitaxy is prevalent in the Sb/graphene system remains open. Secondly however, complicating elucidation of this question, also recently reported "remote" epitaxy needs consideration in which epitaxy is impressed "remotely" between a deposit and its underlying substrate *through* an intermediate 2D layer.[72] In the present work, this would involve interactions between Sb and the underlying Cu catalyst foils impressed through the graphene monolayer.[38] Notably, in this scenario the graphene could also be required to act as a diffusion barrier to prevent chemical reactions between Sb and Cu,[73,74] thus actually actively facilitating the remote epitaxy. Third, in contrast to the atomic-scale epitaxy, the last possibility involves macroscopic corrugations on the support (e.g. Cu surface steps) that result in alignment via preferred heterogeneous nucleation sites (e.g. at steps) and diffusion directing effects.[75]

From SEM data as in Fig. 1a alone, these three possible causes of the observed Sb alignments are hard to disentangle: First, direct vdW epitaxy would be readily compatible with the observed lateral length scales of alignment in Fig. 1a as the lateral size of our CVD graphene domains is in the tens of μm range.[61,62,68] Therefore the field of view in Fig. 1a represents most likely only one single-crystalline graphene domain (although not confirmable by SEM) which could facilitate rotational alignment over the entire field of view. Second however, graphene-mediated remote epitaxy between Sb and Cu is also conceivable for Fig. 1a, since the Cu grain sizes in our Cu foils after graphene CVD are in the mm-range.[61,62] Notably, no Cu grain boundary is visible in Fig. 1a,[62] thus confirming a single Cu orientation across the field of view in Fig. 1a. However, as we show in Supporting Fig. S5, direct deposition of Sb on Cu (i.e. without graphene in between) does not show any indications of epitaxial order in the Sb deposits under our conditions.[73] Nevertheless, a graphene-mediated remote epitaxy mechanism between Sb and Cu[38] cannot be excluded based on Fig. 1a. The third possibility, i.e. surface corrugations on the Cu support, could also direct the Sb deposits, although not resolvable in Fig 1a.

To disentangle these three possible influences, we investigate in Fig. 3 the relative orientation of Sb deposits at 150 °C and 250 °C *directly* onto suspended monolayer graphene membranes, i.e. *without* Cu foil underneath. In doing so, we exclude any possible indirect influence of Cu underneath the graphene on the Sb alignment (i.e. we exclude "remote" epitaxy and an influence from Cu surface corrugations). In particular, we correlate the STEM-derived orientation (via FT analysis) of the lattice of the Sb deposits (β-Sb(001): Fig. 2a-g; cubic Sb(001): Fig. 2i-n) with the underlying graphene lattice orientation measured adjacent to the Sb deposit within a few nm distance[68] for multiple Sb deposits of both morphologies. Via plotting histograms of the relative rotational (mis)orientations of the graphene [120] orientation and prominent orientations in the two respective Sb lattices (β-Sb [110]: Fig. 2g; cubic Sb [110]: Fig. 2n) we find clear peaks in the (mis)orientation distributions for both Sb phases. This unambiguously suggests direct epitaxy effects to be present between the graphene and *both* Sb phases. In particular 2D β-Sb shows a preferred misorientation of ~0° between the graphene [120] and the β-Sb [110] in-plane directions (i.e. β-Sb [110]||graphene[120]), as shown in the model in Fig. 2e. Additionally, a secondary, less prominently preferred misorientation appears at an offset of ~30° for β-Sb [110] and graphene [120] in Fig. 2e. For cubic Sb we find a single preferred misorientation of ~0° between the graphene [120] and cubic Sb [110] directions (i.e. cubic Sb[110]||Graphene[120]), as depicted in the model in Fig. 2l.

To cross-check these STEM-derived rotational vdW epitaxy relations via the SEM data in Fig. 1a we label the salient directions by colored arrows in Fig. 1a: We first assign the long axis of the cubic Sb rods to cubic Sb [110] based on Fig. 1e-g (red arrow). Based on cubic Sb[110]||Graphene[120], this direction then coincides with graphene [120] direction (brown arrow). Thereby it becomes apparent that for most of the β-Sb triangles in Fig. 1a one triangle edge (red arrow) coincides with the graphene[120] direction. This is exactly as expected from the β-Sb [110]||graphene[120] relation and from the observation that one edge direction of the triangles is typically β-Sb [110], as inferred in Fig. 1c,d. Consequently, STEM and SEM data consistently suggest *direct* Sb/graphene vdW epitaxy for both β-Sb and cubic Sb with the preferred overall relations β-Sb(001)||graphene(001)/β-Sb[110]||graphene[120] and cubic Sb(001)||graphene(001)/cubic Sb[110]||graphene[120], respectively.

Prior work has investigated possible epitaxy between Sb polymorphs and graphene (and graphite) with mixed results: Early work did not find evidence for epitaxy in β-Sb/graphite (but had only limited statistics measured).[51] Also recent other studies did not observe epitaxy in β-Sb/graphene.[31,44] In contrast, another recent study of β-Sb and Cu-supported graphene suggested epitaxy for β-Sb/graphene to exist based on indirect measurements, identifying two preferred orientations of (in our notation) 0° and 30° offset between β-Sb [110] and graphene [120].[38] This is in good agreement with our findings in Fig. 3g which are based on direct observations of the β-Sb/graphene interface. For the β-Sb[2-21]/cubic Sb(001)/graphite system no evidence for epitaxy has been reported prior.[31,51] In contrast, we here find strong evidence also for rotational vdW epitaxy in the β-Sb[2-21]/cubic Sb(001)/graphene system. Combined, our observations show that vdW epitaxy can be enforced on 2D and 1D Sb deposits on graphene.

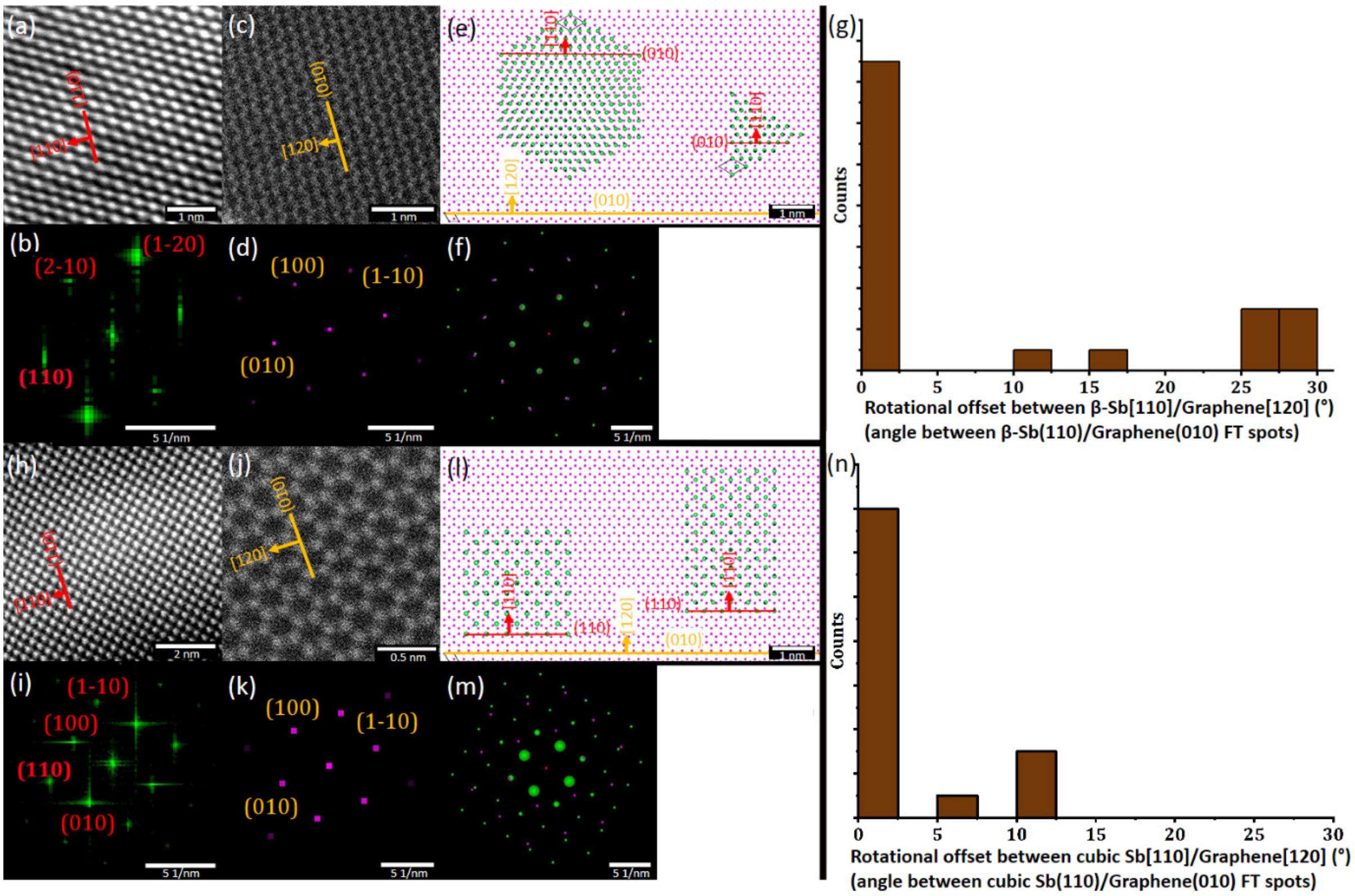

**Figure 3:** (a) Atomic resolution ADF STEM and (b) corresponding FT of a β-Sb(001) particle on suspended graphene. (c) ADF STEM and (d) FT of the graphene(001) lattice just adjacent to the particle in (a). The FTs are indexed to their corresponding phases and in the ADF images salient planes and directions are highlighted. (e) Atomic models showing in-plane vdW epitaxial relations derived from the data in (a-d), suggesting β-Sb(001)||graphene(001)/β-Sb[110]||graphene[120] i.e. angle between β-Sb[110] and graphene[120] = 0°. (f) shows the overlay of simulated FTs corresponding to the models in (e). (g) Histogram of multiple measurements similar to (a-f) showing a distribution of (mis)rotation angles between β-Sb[110] and graphene[120] which peaks at 0°, confirming the suggested in-plane vdW epitaxy relation depicted in (e) to be preferred. (h-n) Corresponding measurements for cubic(001) deposits on suspended graphene, yielding a preferred cubic Sb(001)||graphene(001)/cubic Sb[110]||graphene[120] in-plane vdW epitaxy relation.

**Oxidation Susceptibility.** After having identified the nature of our Sb deposits and their relation to the Graphene support, we turn to the oxidation susceptibility of our Sb deposits. Oxidation susceptibility is of significant importance in terms of processing and applications. Additionally, 2D Sb-oxides are beginning to attract research interest in their own right.[64,76,77] While Raman spectroscopy in Fig. 1b did not suggest significant Sb-oxide presence in our samples, close inspection of the β-Sb(001) FT in Fig. 1d reveals another, weaker intensity set of spots of six-fold symmetry at lower k-vectors (indexed "*" at ~0.4 nm) than the six-fold (110) β-Sb(001) spot family which is indexed in the FT. These weak inner spots may be identified with the presence of cubic $Sb_2O_3$ viewed along the [111] zone axis[64] (i.e. cubic $Sb_2O_3$(111), see Supporting Fig. S1 and S2). This poses the question whether our β-Sb deposits are partly and/or superficially oxidized during sample storage in ambient air. Some prior work has reported stability of antimonene against oxidation in the ambient conditions[33,35,36,42,43,46,78] but other work has suggested thin antimony oxide present around Sb structures to be also prevalent.[22,23,26,38,79–81] Notably, for the β-Sb[2-21]/cubic Sb(001) deposits no signs of additional *crystalline* oxides are found in (S)TEM or FT data.

To investigate possible oxidation effects for our Sb deposits in a localized fashion we use chemical identification via electron energy loss spectroscopy (EELS). In EELS of Sb/Sb-oxide mixtures, compositional analysis based on the commonly used EELS core loss regions is however difficult since the core loss Sb $M_{4,5}$ edge at ~528 eV (which follows a delayed maxima fashion) is very close/partially overlapping the O K-edge at ~532 eV (Supporting Fig. S6).[82,83] An alternative approach is investigating the valence EELS (VEELS) low loss region in which a sharp bulk plasmon peak at ~16.8 eV is related to metallic Sb,[84] while the plasmon peak shifts for Sb-oxides to a distinctly higher energy of ~22eV.[82] For VEELS, however, also the graphene support (and, if present, amorphous carbon TEM grid membrane) has to be considered with plasmon signatures at ~27 eV.[85]

In Fig. 4a a typical VEELS spectrum acquired from a flat 2D β-Sb crystal (inset) is shown. We note that this particular Sb deposit was characterized by VEELS after ~8 months of ambient air storage, thus allowing us to probe long-term resilience against oxidation. We find that a sharp metallic Sb VEELS peak is dominating the fitted VEELS spectrum with only a small contribution of the Sb-oxide component present even after the long term air exposure. VEELS data for β-Sb[2-21]/cubic Sb(001) deposits shows similar results. The VEELS findings are thereby in agreement with the Raman data in Fig. 1b that suggested metallic Sb to be dominant in our deposits. The VEELS results however suggest the possibility of a very thin superficial oxide layer (which might be below the detection limit for Raman). This further implies for the β-Sb(001) that the inner reflections (labelled "*") in the FT in Fig. 1d may be indeed related to a very thin crystalline cubic $Sb_2O_3$(111) overlayer on the elemental 2D β-Sb(001) crystal (where reflection "*" in Fig. 1d corresponds to the (2-20) reflection in $Sb_2O_3$, see also Supporting Figs. S1 and S2). This suggests the possibility of intrinsic Sb-oxide/Sb/graphene heterostructure formation from simple ambient air exposure. In particular, whenever present, the six-fold $Sb_2O_3$ (2-20) reflection family consistently has a rotational misorientation of ~30 ° with the six-fold β-Sb (110) reflection family (as in Fig. 1d). This indicates an epitaxial relation $Sb_2O_3$(111)||β-Sb(001)/$Sb_2O_3$[2-20]||β-Sb[110] for the Sb-oxide/Sb interface. See Fig. 4b

for an atomic model of the suggested Sb-oxide/Sb heterostructure. Unfortunately, the top regions in the β-Sb flakes in our cross-section (S)TEM data are all not well enough resolved (due to Pt/C protection layers) to finally fully confirm the suggested presence of this ultrathin epitaxial $Sb_2O_3$ overlayer. We note however that prior x-ray photoelectron spectroscopy measurements of 2D Sb oxidation found oxide stoichiometries consistent with the here suggested crystalline $Sb_2O_3$ phase,[80] and that our core loss EELS in Supporting Fig. S6 also is best matched with $Sb_2O_3$ stoichiometry. No such crystalline overlayers are suggested from our data for the β-Sb[2-21]/cubic Sb(001), albeit an amorphous Sb-oxide overlayer could be present. Combined, our microscopic and spectroscopic data shows that while our Sb deposits are overall well resilient against environmental bulk oxidation, the possibility of superficial oxidation in ambient air still requires consideration, in particular since for β-Sb(001) deposits the formation of an epitaxial $Sb_2O_3$(111) oxide overlayer is inferred from our data.

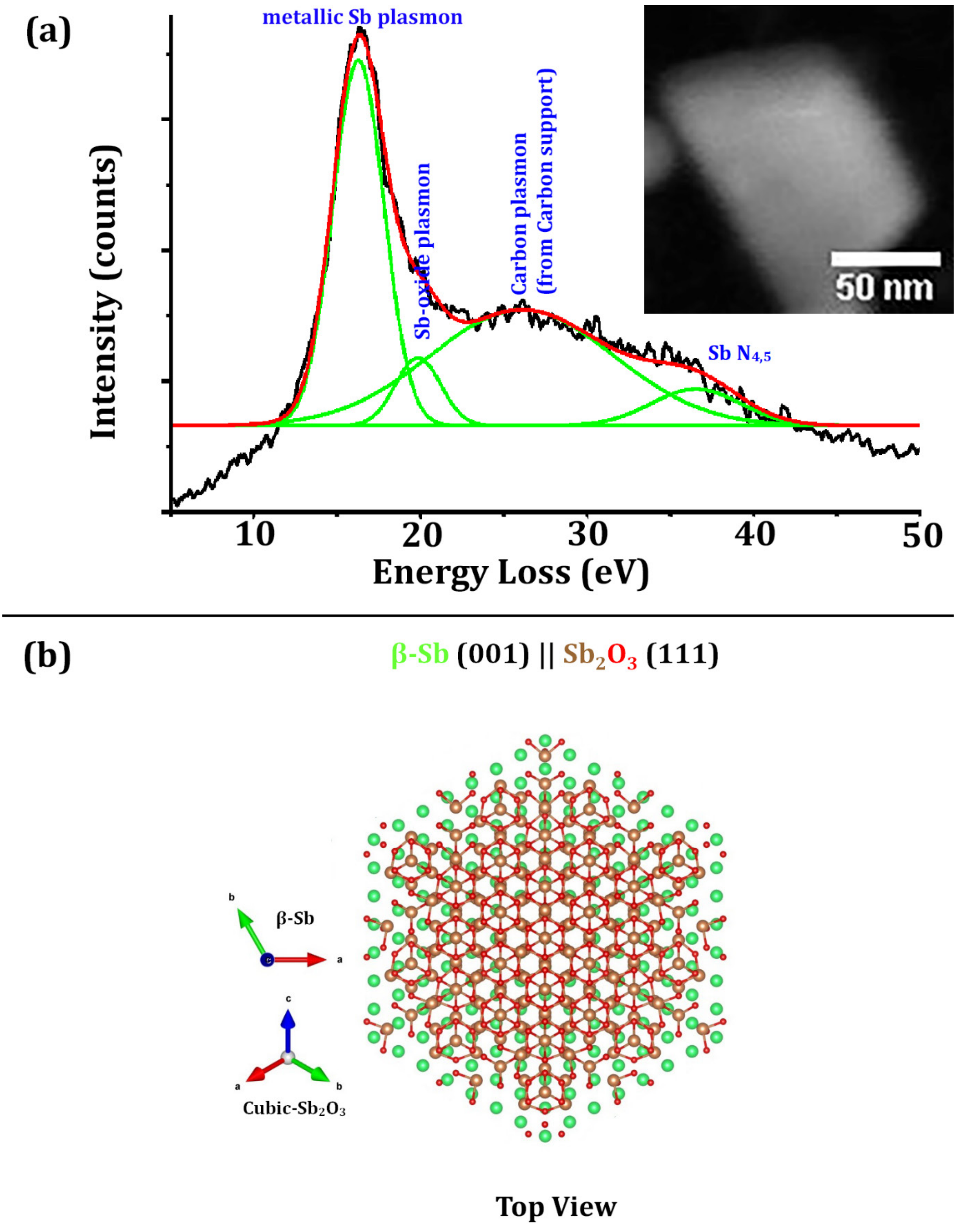

**Figure 4:** VEELS spectrum of the β-Sb(001) crystal on suspended graphene in the ADF STEM in the inset. The spectrum was acquired after ~8 months ambient air exposure of the sample. The VEELS data is fitted to the components labelled and described in the main text. (b) Atomic model of the suggested the $Sb_2O_3$(111)||β-Sb(001) heterostructure that forms from ambient air exposure on β-Sb(001) crystals.

**Growth Parameter Space.** Finally, we examine the wider parameter space of our Sb PVD. Fig. 5 compares Sb deposition results of nominally 10 nm Sb (regulated via a co-exposed (non-heated) quartz crystal microbalance) as function of substrate temperature from RT to 250 °C onto CVD graphene-covered Cu foils (Fig. 5a,c,e, SEM) as well as directly onto suspended graphene membranes (no Cu underneath, Fig. 5b,d,f, TEM). Fig. 5 shows that the Sb deposit morphology drastically changes between RT and elevated temperature (150 °C, 250 °C) depositions: For RT depositions (Fig. 5a,b) merged (truncated) (semi-)spherical features dominate. High resolution STEM in Supporting Fig. S7 shows that these RT-deposited (truncated) (semi-)spheres are amorphous. For 150 °C depositions (Fig. 5c,d) the above described triangular/hexagonal shaped 2D β-Sb and rod-shaped cubic Sb crystals along with few (semi-)spherical Sb deposits are found. Among the 2D β-Sb deposits the hexagonal base shape is more prevalent. For 250 °C depositions (Fig. 5e,f) practically only triangular/hexagonal shaped 2D β-Sb and rod-shaped cubic Sb are found, whereby now among the 2D β-Sb triangles dominate. Notably, for neither 150 °C nor 250 °C we find evidence for an underlying continuous Sb layer on neither Cu-supported nor freestanding graphene, the former in contrast to prior literature.[38] Besides deposit morphology, also coverage and retained Sb amount of the nominally 10 nm Sb deposits is strongly influenced by substrate temperature during Sb deposition and, notably, also strongly dependent on substrate-type. In particular, Sb coverage and retained Sb amount strongly decrease with increasing substrate temperature. For RT depositions a homogeneous coverage close to 100 % is achieved on both Cu-supported and freestanding Graphene in Figs. 5a,b and for RT samples good agreement between nominal 10 nm thickness and AFM-calibrated average deposit thickness was found. In comparison, the coverage for 150 °C and 250 °C depositions decreases, whereby the coverage decrease with substrate temperature is even more prominent on the freestanding Graphene (150 °C~40 %, 250 °C: <5 %) than on the Cu-supported Graphene (150 °C: ~40%, 250 °C: ~20%). For Sb deposits with an average thickness of 21 ± 14 nm at 250 °C (see above) this equates to a reduction in Sb amount deposited from RT to 250 °C of ~50 % on Cu-supported Graphene and of ~90 % on freestanding Graphene, respectively. Notably, also the size of individual deposits of Cu is significantly larger than on the suspended graphene, best seen in the 250 °C depositions (Fig. 5e,f). Combined, this suggests a key influence of temperature dependent desorption processes on Sb nanostructure growth.[35,37,47–51] In particular, the balance of Sb adsorption flux ($F_{Sb,ad}$) from the evaporation source onto the graphene substrate and a substrate-temperature-dependent Sb desorption flux from the graphene substrate into vacuum ($F_{Sb,de}(T)$) is key: The observed Sb morphologies imply that at RT $F_{Sb,ad} >> F_{Sb,de}(RT)$ resulting in strong deposition, while the low temperature hinders crystallization of the resulting Sb deposits (possibly via incomplete fragmentation of physisorbed $Sb_4$ species which are the preferred arriving Sb vapor species[51]). This leads to the observed fully covering amorphous Sb deposits at RT. Increasing substrate temperature leads to a strong increase in $F_{Sb,de}(150\text{-}250\ °C)$, reducing the net retained amount of Sb at higher substrate temperature. In turn the higher substrate temperatures facilitate crystallization of the retained Sb deposits (possibly via thermally activated fragmentation of surface-bound Sb species[51] and thus increased Sb re-arrangement). Thereby we grow crystalline Sb deposits with an onset temperature of crystallization of ~150 °C. The observation that this temperature dependence is more pronounced on freestanding graphene membranes as compared to Cu-foil supported graphene, we suggest to be related either to intrinsic substrate effects whereby the Cu surface states underneath the Graphene modify e.g. sticking coefficients to the Sb flux (akin to Cu supports modifying the surface properties of graphene in liquid wetting[86,87]) or to a different local temperature profile on Cu-foil supported graphene vs. suspended graphene membranes due to the macroscopic thickness 25 µm Cu foils. In the latter scenario, the Cu foil acts as an effective heat sink for the additional energy arriving with the incoming Sb flux $F_{Sb,ad}$ compared to the vacuum-suspended monolayer graphene membranes (and, if present, thin amorphous carbon support), thus resulting in a (slightly) lower actual local substrate temperatures on the Cu supported graphene.

The here observed temperature dependence of Sb deposit morphology, crystallization onset and retained Sb amount is in good agreement with prior literature.[35,37,47–51] Beyond this, our results confirm that not only the directly supporting growth substrate (here, monolayer graphene) but also the supporting material *underneath* (here, Cu vs. vacuum) can strongly influence Sb nanostructure growth results.[38] This is important to consider when designing Sb 2D/2D heterostructure stacks. Finally, in Fig. 5g,h we show that the here derived understanding of the balance of adsorption, nucleation, desorption and "sub-support" can also be advantageously employed to engineer larger Sb deposits of high crystalline quality. Fig. 5g,h shows deposition of nominally ~50 nm Sb at 250 °C on Cu-supported (Fig. 5g) and freely suspended graphene (Fig. 5h). On the Cu-supported graphene increasing the deposited Sb amount led not to laterally larger Sb domains but to the onset of undesired three-dimensional Sb overgrowth (Fig. 5g). In contrast, the relatively higher desorption on the suspended graphene enabled a lower Sb nucleation density and consequently a desired larger lateral growth of remaining Sb crystallites (Fig. 5h, since presumably desorption probability decreases with increasing deposit radius). Thereby by adjusting the substrate *underneath* the actual graphene support, we obtain a lateral size of >400 nm diagonal for β-Sb(001) and >500 nm long axis for β-Sb[2-21]/cubic Sb(001) deposits, respectively. This is an improvement not only over the undesired three-dimensional Sb overgrowth from Cu/graphene supported 250 °C/50 nm but also an improvement of factor ~2 compared to the 250 °C/10 nm Cu/graphene-supported deposits. This introduces the substrate *underneath* the direct 2D support as an important parameter to consider in 2D Sb deposition.

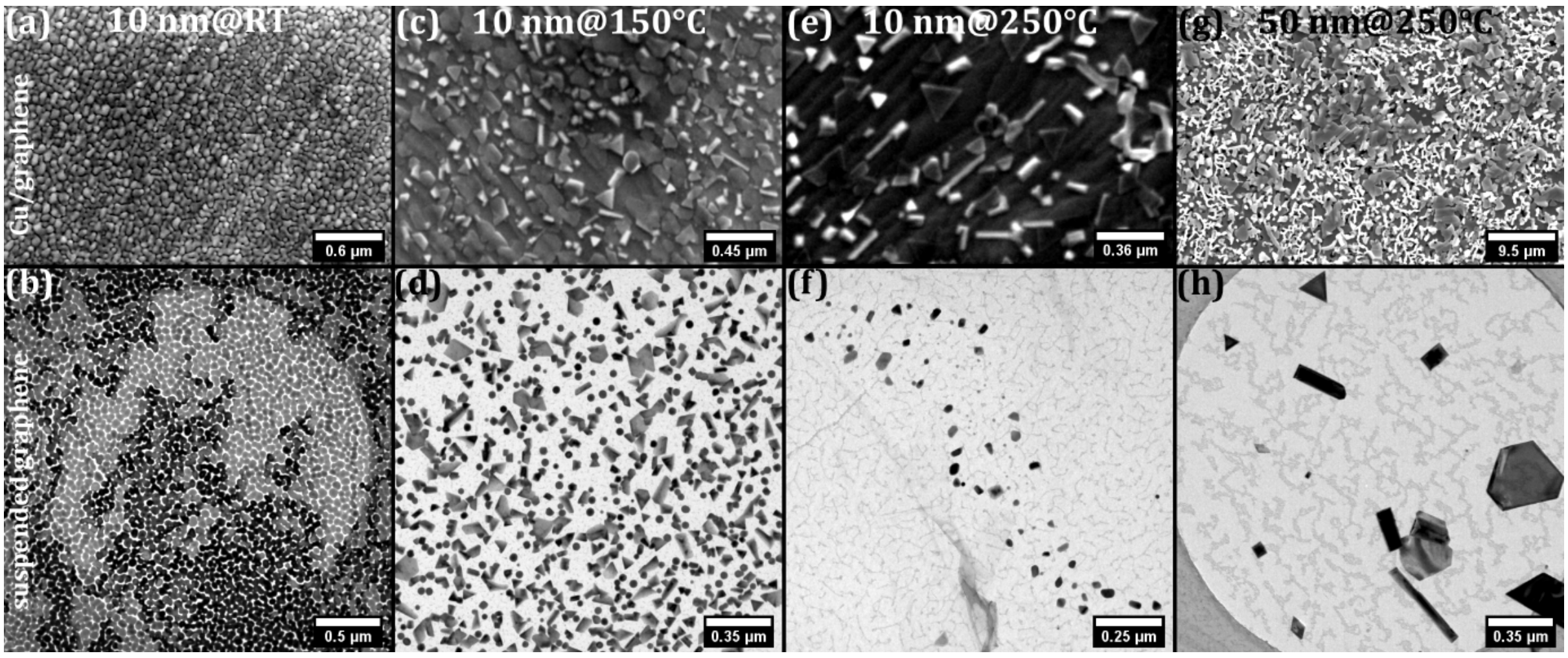

**Figure 5:** Sb deposits on Cu-supported graphene (SEM in a,c,e,g) and suspended graphene (BF TEM in b,d,f,h) deposited at nominal thicknesses of 10 nm at RT (a,b), 150 °C (c,d) and 250 °C (e,f) and 50 nm at 250 °C (g,h), respectively.

## Conclusions

In summary, using high resolution STEM, we elucidate the structural relations in 2D Sb/graphene heterostructures which we present as a model system for 2D Sb's use in electronics and energy applications. We find two Sb morphologies to co-exist under optimized deposition conditions: Few-layer 2D β-Sb(001)||graphene(001) and 1D Sb which can be matched to both Sb[2-21]⊥graphene(001) and cubic Sb(001)||graphene(001). Notably, both morphologies exhibit direct in-plane rotational vdW epitaxy with the graphene support. Both morphologies are stable against ambient air oxidation even for prolonged storage, albeit superficial surface Sb-oxide formation is found. Notably, for β-Sb(001) growth of an epitaxial $Sb_2O_3$(111)||β-Sb(001) overlayer is suggested from our data. While exact Sb growth results depend on growth parameters such as temperature, notably also the nature of the support *under* the direct graphene support is found to have a key influence on Sb growth. Combined, our findings explore at high resolution the structural diversity in scalably fabricated 2D Sb and in 2D Sb/graphene heterostructures.

**Acknowledgements:**

B.C.B, K.E., S.H. and C.M. acknowledge support from the Austrian Research Promotion Agency (FFG) under project 860382-VISION. We also acknowledge use of the facilities at the University Service Centre for Transmission Electron Microscopy (USTEM), Vienna University of Technology (TU Wien), Austria for parts of this work.

**Notes:**

At the time of performing of this study S.H has been affiliated with GETec Microscopy GmbH., Austria.

Supporting Information to:

# Resolving Few-Layer Antimonene/Graphene Heterostructures

Tushar Gupta,[1] Kenan Elibol,[2] Stefan Hummel,[2,3] Michael Stöger-Pollach,[4] Clemens Mangler,[2] Gerlinde Habler,[5] Jannik C. Meyer,[2] Dominik Eder,[1,*] Bernhard C. Bayer[1, 2,*]

*1. Institute of Materials Chemistry, Vienna University of Technology (TU Wien), Getreidemarkt 9/165, A-1060 Vienna, Austria.*

*2. Faculty of Physics, University of Vienna, Boltzmanngasse 5, A-1090 Vienna, Austria*

*3. GETec Microscopy GmbH, Seestadtstrasse 27, A-1220 Vienna, Austria*

*4. USTEM, Vienna University of Technology (TU Wien), Wiedner Hauptstrasse 8-10, A-1040 Vienna, Austria*

*5. Department of Lithospheric Research, University of Vienna, Althanstrasse 14, A-1090 Vienna, Austria*

**Corresponding authors: bernhard.bayer-skoff@tuwien.ac.at, dominik.eder@tuwien.ac.at*

## Materials and Methods

Physical vapor deposition (PVD) of Sb employed a commercial thermal evaporation system (MANTIS deposition system QUBE series) with a base pressure of $4\times10^{-5}$ mbar. For PVD Sb powder (Goodfellow, 99.999% purity, average particle size 150 µm) was loaded into a W boat, which was heated resistively to sublime the Sb. Phase diagrams[1] of W and Sb were cross-checked to ensure that no undesired intermetallics are formed during evaporation. Samples were loaded upside down over the evaporation source and behind a manual shutter. The sample table was electrically heated to a desired substrate temperature, where room temperature (RT, i.e. non heated), 150 °C and 250 °C were employed in this study. The Sb evaporation flux and nominally deposited thickness were monitored *in situ* using a non-heated quartz micro balance (QMB). The nominal Sb thickness QMB measurement was calibrated by evaporation of selected Sb films over partially masked Si wafers ("thickness monitors") at room temperature to measure Sb film thickness over film edges by atomic force microscopy (AFM). Note that the nominal thicknesses quoted in this study refer to the measured thickness values obtained from the non-heated QMB and from these Si wafer calibration depositions at RT. As discussed in the exploration of the parameter space of Sb PVD in the main text, actual retained Sb thicknesses may strongly reduce as a function of increasing substrate temperature and also type of substrate type via desorption effects.

Substrates for Sb deposition were chemical vapor deposited (CVD, 960 °C in $CH_4/H_2/Ar$ at ~12 mbar) polycrystalline (grain size tens of µm) monolayer graphene films remaining on their 25 µm thick Cu foil catalysts[2,3] as well as CVD graphene films suspended as freestanding monolayer membranes over the regular hole arrays in an amorphous carbon film of a transmission electron microscopy (TEM) grid (Quantifoil) i.e. no Cu underneath.[4] For graphene-free reference also Cu foils without graphene were prepared as substrates by annealing at 960 °C in 12 mbar $H_2$ without $CH_4$.

The scanning electron microscopy (SEM) employed a FEI Quanta 250 FEG SEM. TEM studies incl. bright field (BF) imaging, selected area electron diffraction (SAED), (valence) electron energy loss spectroscopy ((V)EELS) and energy dispersive X-ray spectroscopy (EDX, confirming the Sb purity) were performed on a FEI TECNAI F20 at 60 kV electron acceleration voltage. Scanning transmission electron microscopy (STEM) studies were performed in an aberration-corrected NION ULTRA STEM100 at 60 kV electron acceleration voltage and in (high angle) annular dark field ((HA)ADF) mode (80 to 200 mrad).[5] Correlative AFM-SEM studies employed a GETec AFSEM module installed in a FEI Quanta 600F SEM. Conventional AFM studies employed a NT MDT NTEGRA Spectra in tapping mode. AFM analysis employed Gwyddion software.[6] Raman spectroscopy employed a Horiba LabRAM at 532 nm laser excitation wavelength. Cross-sections for TEM/STEM of Cu/graphene/Sb stacks were cut by focused ion beam (FIB) processing in a FEI Quanta 3D FEG. A protective C and Pt bilayer was deposited locally by onto the region of interest prior to FIB cutting.

Phase analysis of (S)TEM data employed primarily Fourier Transform (FT)/SAED pattern simulation using Highscore Plus/Pdf4+ software (ICDD Pdf4+ 2020 RDB: Software version: 4.20.0.1. Database version: 4.2001.) for manual matching of measured and simulated FT/SAED patterns. Additionally, also automated phase identification of measured FT/SAED data was performed using JEMS software. Structure visualization was done by Vesta[7] software. In particular the following structural database entries were found to best fit our measured FT and SAED (Pdf4+ code/Inorganic Crystal Structure Database ICSD collection code/literature reference): β-Sb: 04-14-2871/55402/ref. [8]; simple cubic Sb: 04-13-3319/651499/ref. [9]; $Sb_2O_3$: 00-042-1466/1944/ref. [10]. Notably, we checked additionally 45 other Sb and 61 other Sb-oxide entries from the ICDD Pdf4+ database which consistently gave worse matches to experimental data.

Note that β-Sb (A7, R-3m, 166) is often described in literature with hexagonal axis (as here) but also with rhombohedral axis.[11–13] Therefore, numerical (hkl) and [uvw] values need consideration of selected hexagonal or rhombohedral axis system, when comparing between reports. Likewise, within the hexagonal axis system some literature uses a *a,b* base vector inner angle of 120 ° (as here) while other literature uses *a,b* base vector inner angle of 60°.[11–14] Again therefore comparison of numerical (hkl) and [uvw] values must consider the selected axis system. To avoid ambiguity the here used axis are typically plotted alongside the atomic models throughout the manuscript.

We calculated the average equivalent feature sizes (see main text) for the Sb deposits from their base areas (computed by ImageJ[15]) in SEM and AFM data as follows: Due to the different asymmetries between the characteristic base shapes for 2D triangular/hexagonal β-Sb(001) and 1D rod-like β-Sb[2-21]/cubic Sb(001) we recalculate a characteristic feature size assuming a square base shape for both phases and defining the side length of this square as the equivalent features size.

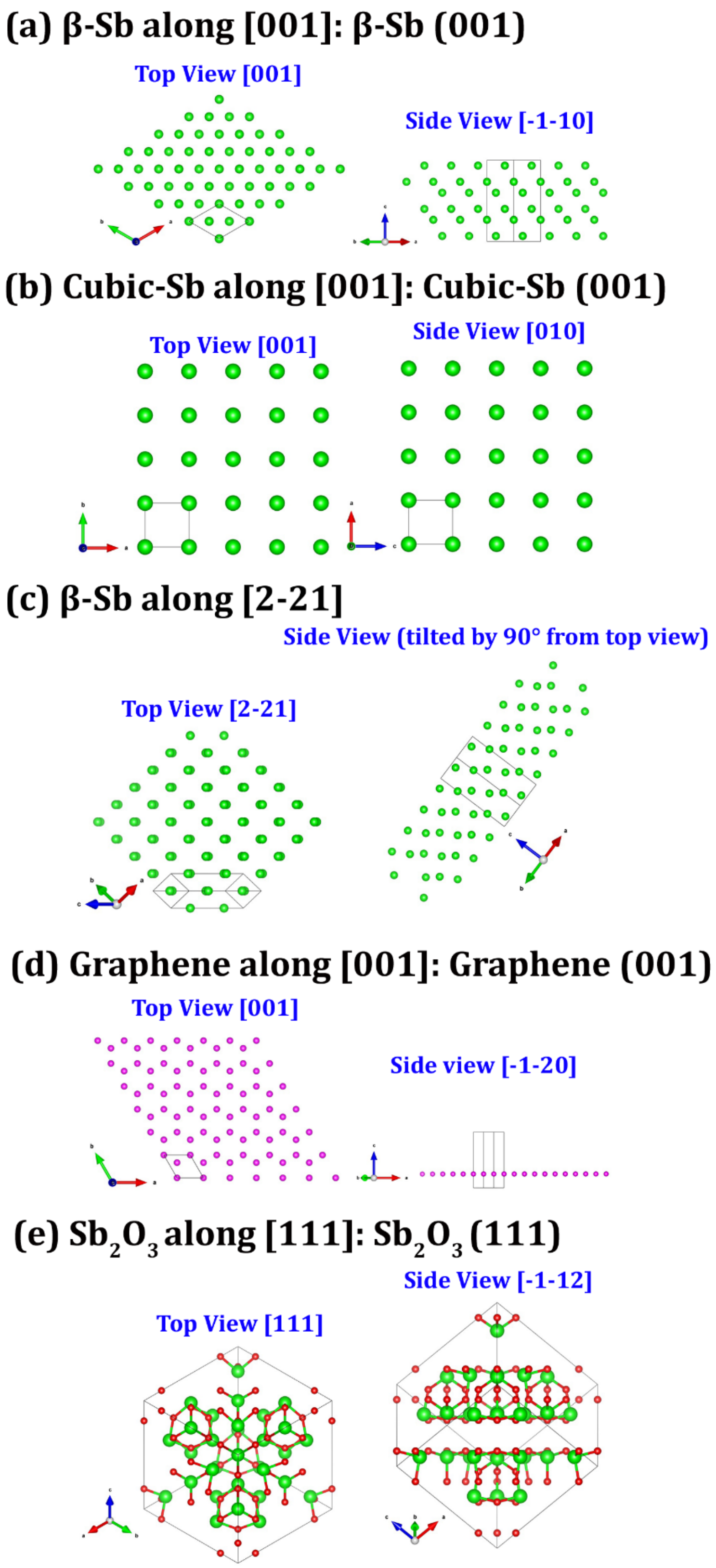

**Supporting Figure S1:** Atomic models in (left) "top/plan view" (as in "normal" ADF STEM and BF TEM of Sb deposits on graphene) and (right) "side" view (as in cross-sectional ADF STEM and BF TEM data) for (a) β-Sb(001), (b) cubic Sb(001), (c) β-Sb[2-21], (d) graphene(001) and (e) cubic $Sb_2O_3$(111). Unit cells and axes are plotted for all phases. The zone axis for all views are indicated. Note that β-Sb with [2-21] zone axis perpendicular to support does not have a defined low (hkl) value interface plane parallel to the support when viewed from the side, but only slightly inclined base planes (c). An approximation for an interface plane is β-Sb(10 -10 23). For corresponding simulated SAED/FT patterns see Supporting Fig. S2.

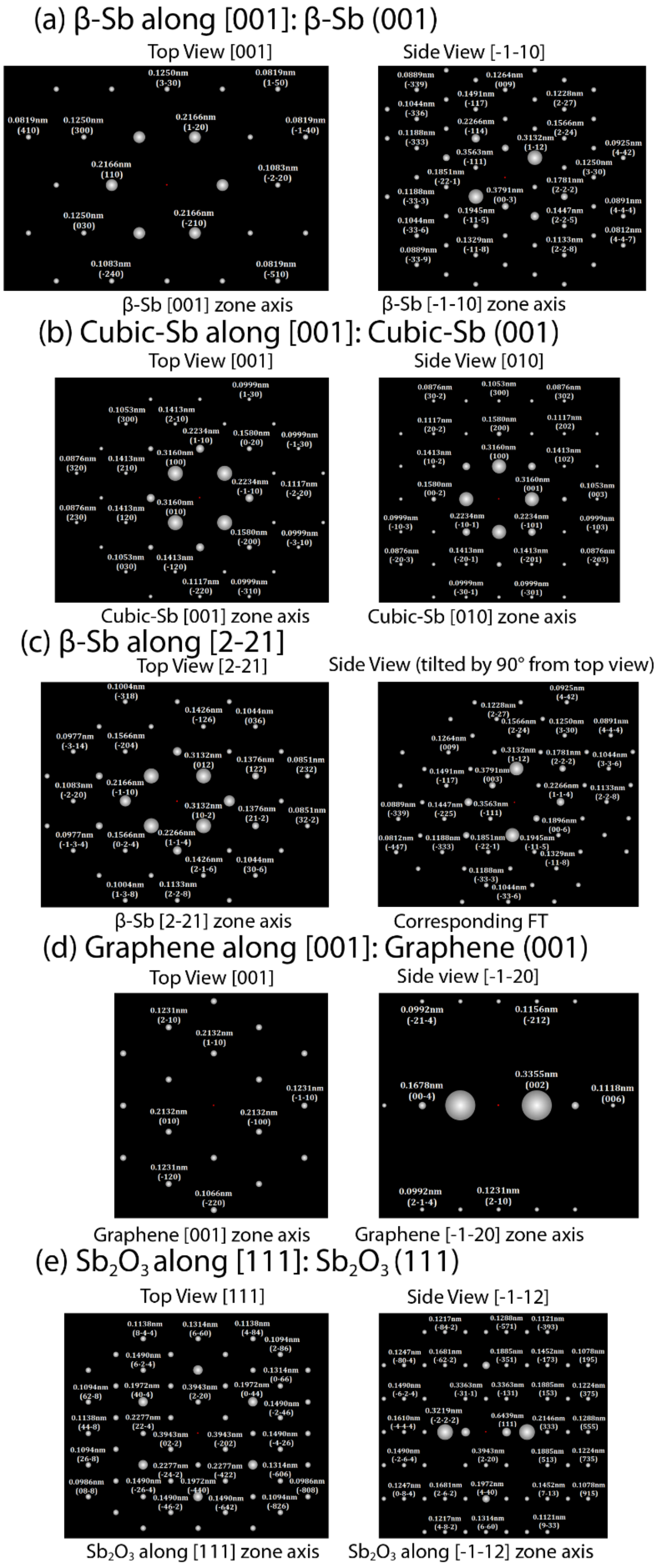

**Supporting Figure S2:** Simulated SAED/FT patterns corresponding to the atomic models in Supporting Fig. S1 from (left) "top/plan view" (as in "normal" ADF STEM and BF TEM of Sb deposits on graphene) and (right) "side" view (as in cross-sectional ADF STEM and BF TEM data) for (a) β-Sb(001), (b) cubic Sb(001), (c) β-Sb[2-21], (d) graphene(001) and (e) cubic $Sb_2O_3$(111). Note that atomic models in Supporting Fig. S1 and simulated SAED/FT patterns here are not rotation corrected and therefore may include arbitrary rotations around the zone axis with respect to each other.

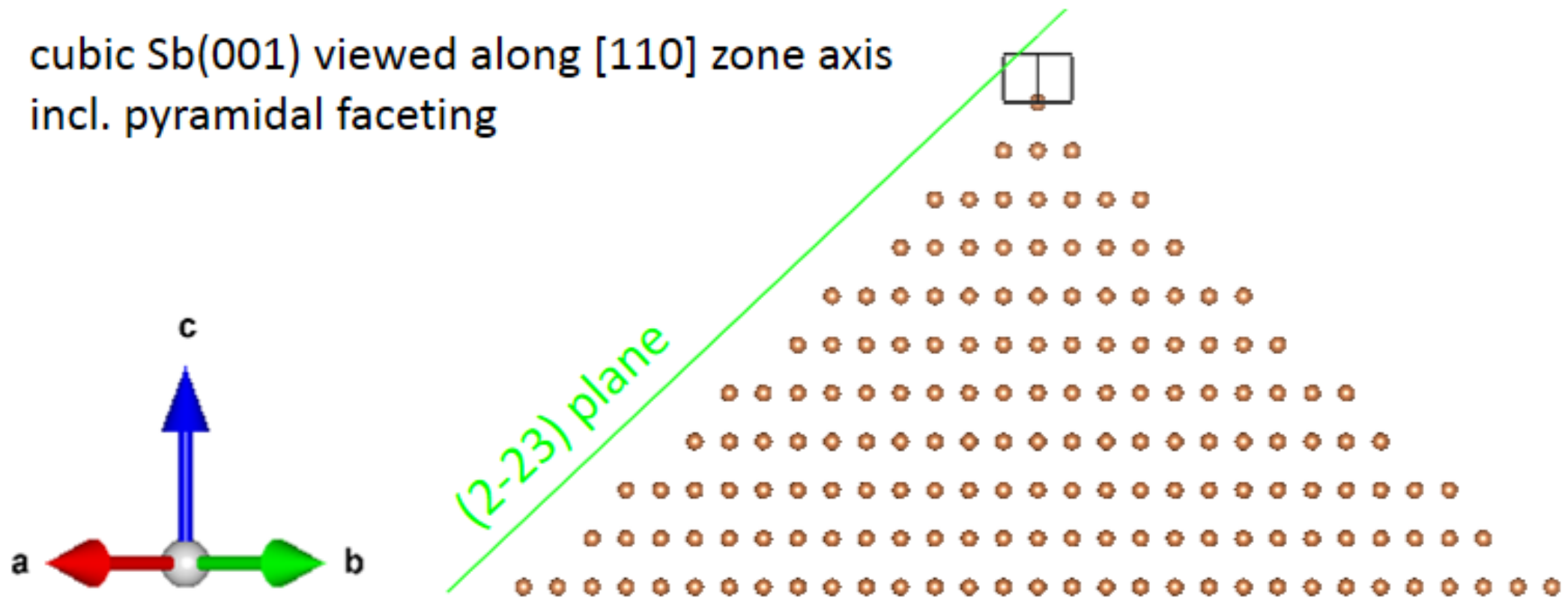

**Supporting Figure S3:** Atomic model of cubic Sb(001) viewed along the [110] zone (i.e. in side view corresponding to Fig. 2e), showing that pyramidal faceting matches with (2-23) plane family delineating surface planes matches the measured angles in Fig. 2e. Consistently, top view projections of the (2-23) faceting match the top view images in Fig. 1e-g.

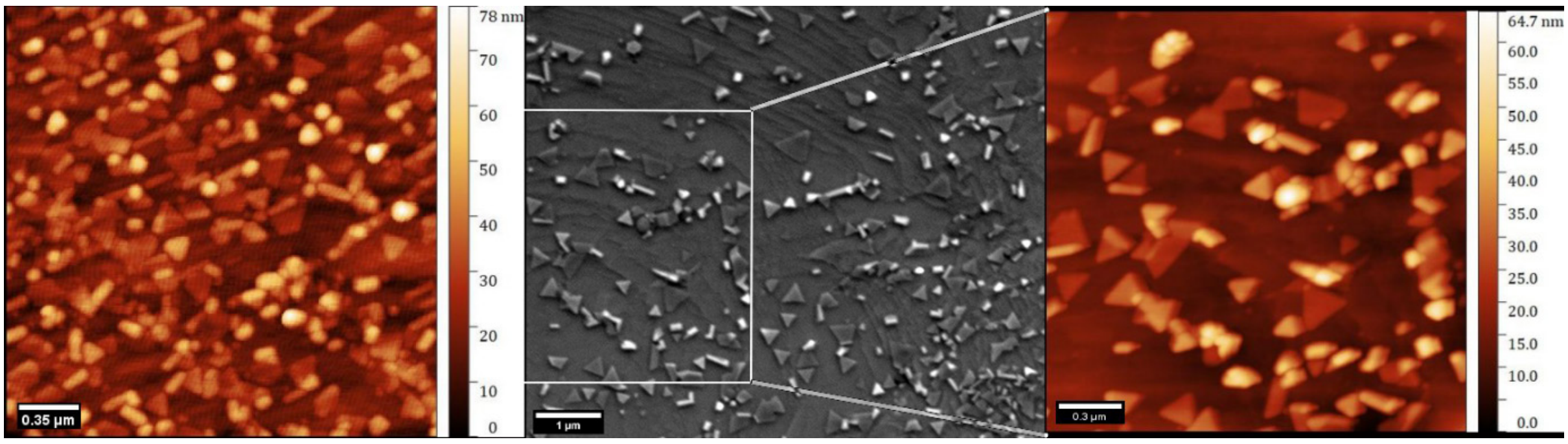

**Supporting Figure S4:** (Left) Conventional AFM of 10 nm Sb deposits on Cu-supported graphene from 250 °C depositions. (Middle and right) Correlated SEM (middle) and AFM (right) from the same sample, measured by a GETec AFSEM.

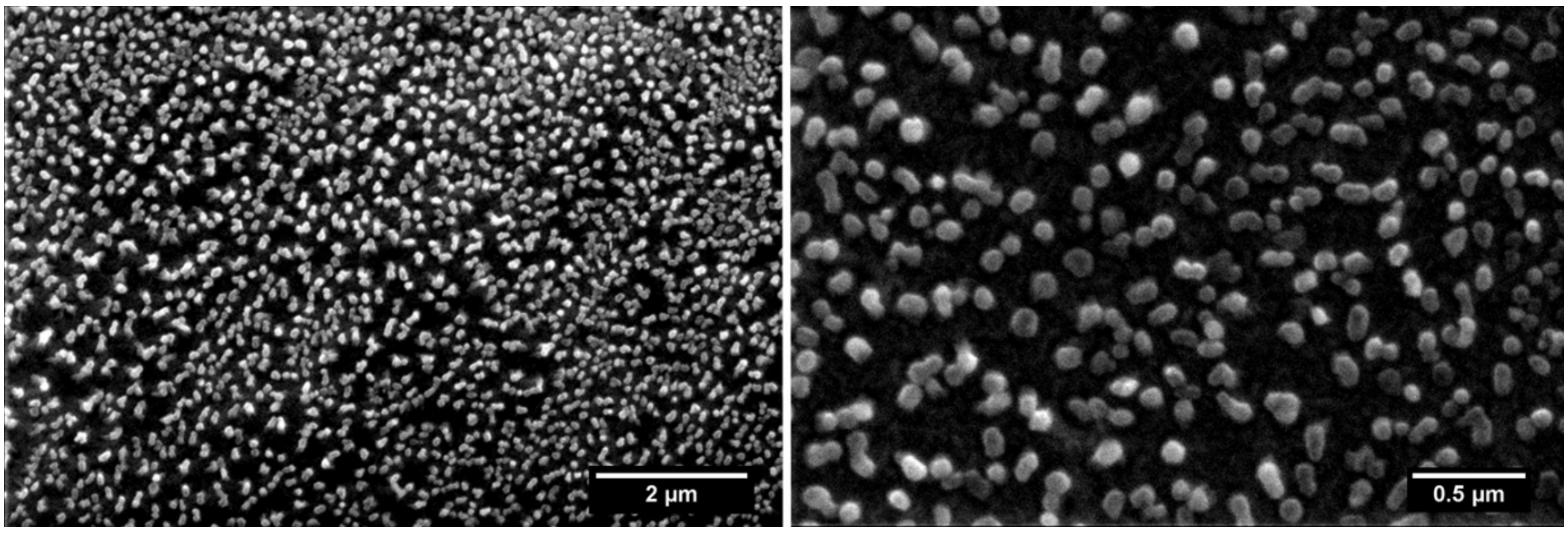

**Supporting Figure S5:** SEM of 10 nm Sb deposition at 250 °C onto bare Cu foils (i.e. *without* graphene). Cu foils have been annealed prior to Sb deposition at 960 °C in $H_2$/Ar (but without $CH_4$ exposure) to obtain a similar Cu grain structure as for the Cu foils after graphene CVD. Notably the Sb deposits on the bare Cu foils do not show the typical hexagonal/triangular and nanorod shapes seen on the Cu-supported graphene, indicating that the monolayer graphene drastically changes Sb growth behavior.

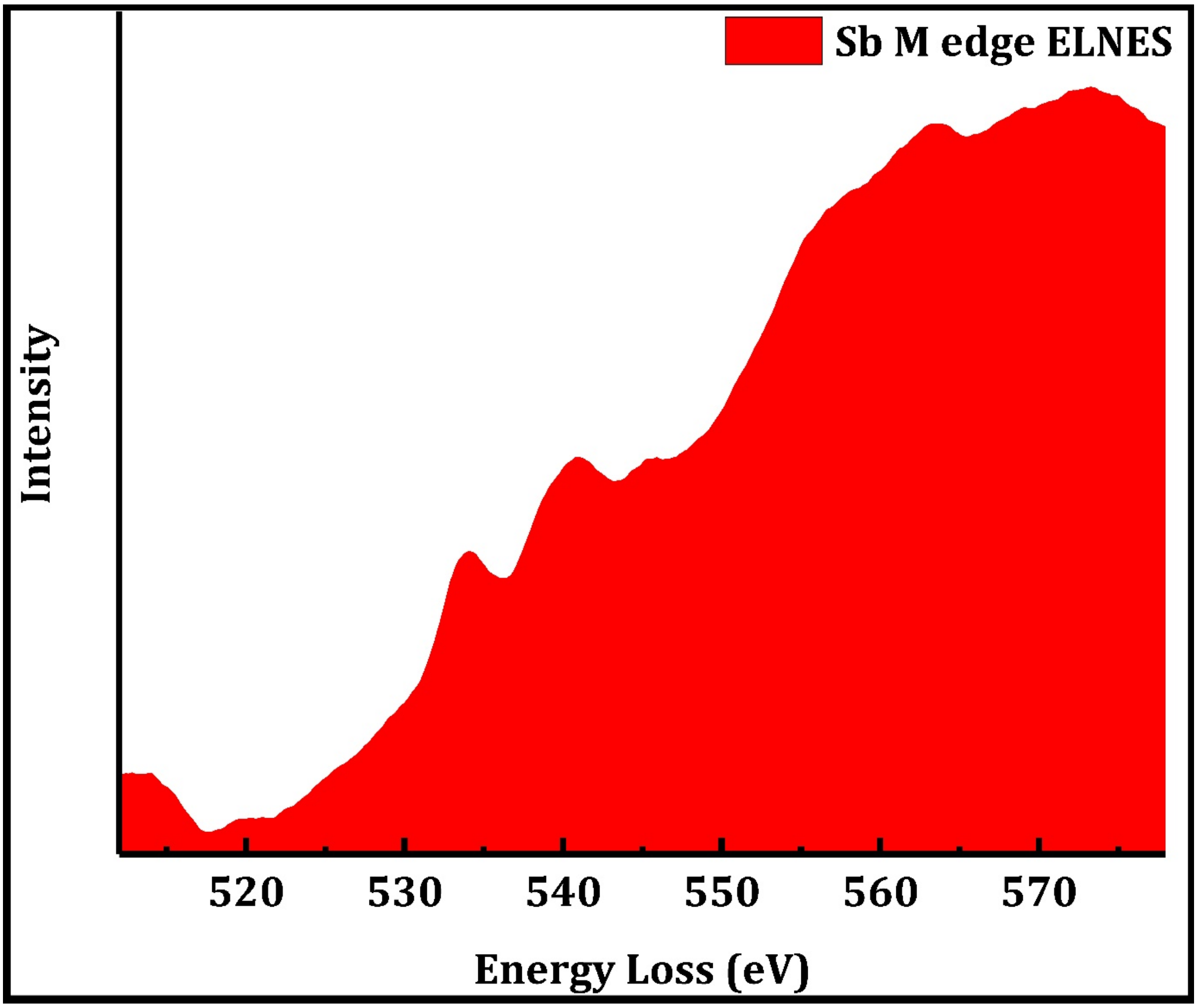

**Supporting Figure S6:** Core loss EELS spectrum of the Sb deposit corresponding to a Sb deposit on graphene as in Fig. 4a. Based on this core-loss EELS spectrum, the presence of some Sb-oxide is indicated (best matched $Sb_2O_3$) but it cannot be confirmed from this spectrum alone whether/how much metallic Sb remains (since metallic Sb/Sb-oxide quantification is not possible from core loss ELLS due to overlap of respective spectral features, see also main text).[16,17]

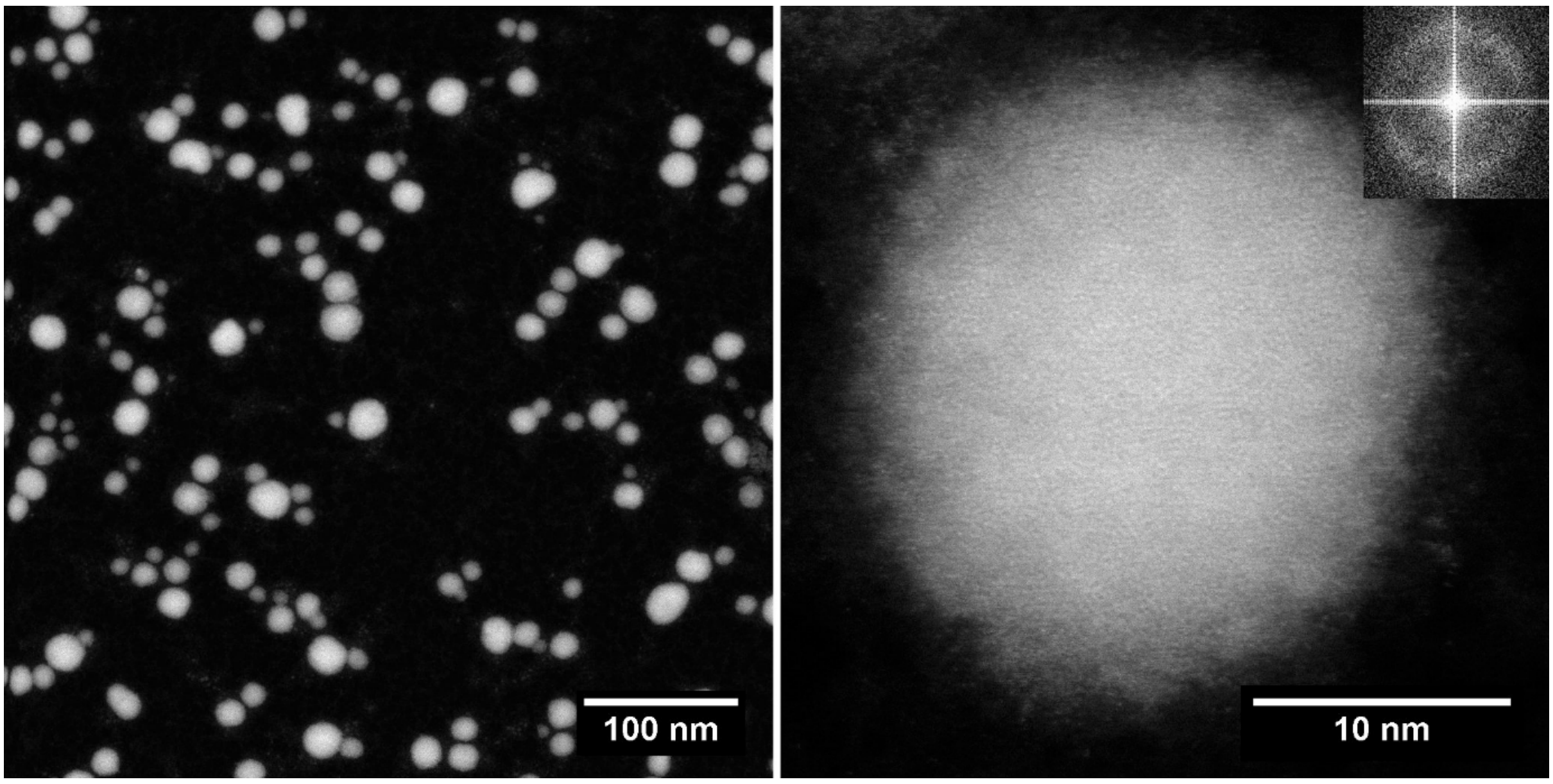

**Supporting Figure S7:** ADF STEM images of RT Sb depositions on suspended graphene at overview (left) and atomic resolution (right). The inset shows the FT to the atomic resolution image. Both visual appearance and FT confirm that the RT deposited Sb on graphene is amorphous. The nominal thickness of the Sb deposits is ~1.6 nm, which was chosen in order to obtain individual particles for STEM imaging.

## Supporting References